# Gravitational lensing by a charged black hole of string theory

A. Bhadra*
*High Energy and Cosmic Ray Research Centre, University of North Bengal, Siliguri 734430, India*

We study gravitational lensing by the Gibbons-Maeda-Garfinkle-Horowitz-Strominger (GMGHS) charged black hole of heterotic string theory and obtain the angular position and magnification of the relativistic images. Modeling the supermassive central object of the galaxy as a GMGHS black hole, we estimate the numerical values of different strong-lensing parameters. We find that there is no significant string effect present in the lensing observables in the strong-gravity scenario.

## I. INTRODUCTION

Superstring theory appears as the most promising avenue to unify gravity with all other fundamental interactions in nature. Although the scattering amplitudes of the graviton in string theory differ considerably from those of pure gravity, still, expanding the string amplitudes about low momentum scales, one can produce an effective action for gravity that includes the Einstein action as the leading term. The correction factors consist of Planck-scale terms in the form of nonvanishing four-loop $\beta$ functions [1]. When the $\beta$ functions are set to zero, the effective field theory admits black hole solutions which can have different qualitative properties from those of general relativity [2]. It is thus of great interest to investigate theoretically the possible stringy effects in a physical observation. But very little has been discussed in the literature in this direction so far. Gegenberg [3] showed that static spherically symmetric solutions of string theory in four dimensions do not lead to string effects in weak-field observations. Rotating charged black hole solutions [4] of heterotic string theory also do not produce any considerable string effect in a Sagnac-like experiment [5]. Thus one has to look for a strong-gravity scenario where a significant contribution of string is expected. The phenomenon of gravitational lensing may provide one such situation, when the lens is a compact object such as a black hole or a neutron star.

Owing to the highly nonlinear character of the Einstein field equations, the theory of gravitational lensing was primarily developed in the weak-field thin-lens approximation [6]. Although these approximations are sufficiently accurate to discuss any physical observation up to today, but with the indication that many nearby galaxies, including our galaxy, host supermassive central black holes [7], the possibility has developed of studying lensing phenomena in the strong-gravity regime which demands a full treatment of lensing theory to any order of approximation. The development of lensing theory in the strong-field regime started recently with the work of Frittelli, Kling, and Newman [8] and that of Virbhadra and Ellis [9]. The former authors provided an exact lens equation without reference to the background space-time. They also noticed that the strong-field thin-lens approximation (without any small angle approximation) can describe lensing satisfactorily even at small impact parameters. On the other hand, Virbhadra and Ellis [9] studied lensing due to the Schwarzschild black hole in an asymptotically flat background by numerical techniques. They showed that while propagating near the black hole horizon light rays take several turns around the lens before reaching the observer, and as a result apart from primary and secondary images a set of infinite images (what the authors termed relativistic images) on both sides of the optic axis will be produced. These relativistic images are the main signature of strong-field lensing but unless the source is almost perfectly aligned with the lens and the observer, these images will be very faint as a result of high demagnification. Bozza *et al.* [10] developed an analytical technique for obtaining the deflection angle in the strong-field situation and showed that the deflection angle diverges logarithmically as light rays approach the photon sphere of a Schwarzschild black hole. The strong lensing due to the Reissner-Nordström (RN) space-time was investigated in [11], and in other work Virbhadra and Ellis [12] studied lensing by naked singularities. Very recently, Bozza [13] extended the analytical theory of strong lensing for a general class of static spherically symmetric metrics and showed that the logarithmic divergence of the deflection angle at the photon sphere is a common feature of such space-times. The Gibbons-Maeda-Garfinkle-Horowitz-Strominger (GMGHS) black hole is a member of this class due to its symmetries, and hence the analytical technique developed in [13] is applicable to this metric.

Following the method of [13], in the present work we wish to study gravitational lensing due to a charged black hole [2] of heterotic string theory with the aim of examining the possible string effects in a strong-field observation. To get a clear idea of the string contribution, we compare the estimated observable lensing quantities due to a charged string black hole with those due to a charged black hole of general relativity.

The paper is organized as follows. The charged black hole solution of string theory is reconsidered in Sec. II, and in Sec. III lensing due to a charged string black hole is presented. By modeling the supermassive central object of the galaxy as a GMGHS black hole, an estimation of observational strong-lensing parameters will be given in Sec. IV along with the similar estimation when the lens is represented by a RN black hole. A discussion of the results is given in Sec. V.

*Email address: aru_bhadra@yahoo.com

## II. THE GMGHS BLACK HOLE

The low energy effective action of the heterotic string theory in four dimensions is given by (in this paper we choose Planck units such that $G=c=\hbar=1$)

$$\mathcal{A}=\int d^4x\sqrt{-g}e^{-\phi}\left(-R+\frac{1}{12}H_{\mu\nu\rho}H^{\mu\nu\rho}-G^{\mu\nu}\partial_\mu\phi\partial_\nu\phi+\frac{1}{8}F_{\mu\nu}F\mu\nu\right), \tag{1}$$

where $R$ is the Ricci scalar, $G_{\mu\nu}$ is the metric that arises naturally in the $\sigma$ model, $F_{\mu\nu}\equiv\partial_\mu A_\nu-\partial_\nu A_\mu$ is the Maxwell field associated with a U(1) subgroup of $E_8\times E_8$, $\phi$ is the dilaton field, and

$$H_{\mu\nu\rho}\equiv\partial_\mu B_{\nu\rho}+\partial_\nu B_{\rho\mu}+\partial_\rho B_{\mu\nu}-[\Omega_3(A)]_{\mu\nu\rho}, \tag{2}$$

where $B_{\mu\nu}$ is the antisymmetric tensor gauge field and

$$[\Omega_3(A)]_{\mu\nu\rho}\equiv\frac{1}{4}(A_\mu F_{\nu\rho}+A_\nu F_{\rho\mu}+A_\rho F_{\mu\nu}) \tag{3}$$

is the gauge Chern-Simons term. It is to be noted that the massless fields arising from compactification have not been included in the effective action. Only a U(1) component of the full set of non-Abelian gauge fields present in the theory has been considered above, and consequently the corresponding solutions carry a U(1) charge only. Assuming $H_{\mu\nu\rho}$ to be zero, the above action in the conformal Einstein frame becomes

$$\mathcal{A}=\int d^4x\sqrt{-g}[-R+2(\nabla\phi)^2+e^{-2\phi}F^2], \tag{4}$$

where the Einstein frame metric $g_{\mu\nu}$ is related to $G_{\mu\nu}$ through the relation

$$g_{\mu\nu}=e^{-\phi}G_{\mu\nu}. \tag{5}$$

The above theory has charged static black hole solutions [2] given by the following field configuration:

$$ds^2=-\left(1-\frac{2M}{r}\right)dt^2+\left(1-\frac{2M}{r}\right)^{-1}dr^2+r^2\left(1-\frac{Q^2e^{-2\phi_0}}{Mr}\right)d\Omega^2, \tag{6}$$

$$e^{-2\phi}=e^{-2\phi_0}\left(1-\frac{Q^2e^{-2\phi_0}}{Mr}\right), \tag{7}$$

and

$$F=Q\sin\theta\, d\theta\wedge d\phi. \tag{8}$$

$\phi_0$ is the asymptotic constant value of the dilaton field. The metric (6), which is often called a Gibbons-Maeda-Garfinkle-Horowitz-Strominger black hole, describes a black hole of mass $M$ and charge $Q$ when the ratio $Q/M$ is small. Otherwise it exhibits a naked singularity. Similar behavior is also shown by the RN solution of the Einstein-Maxwell theory, which is given by

$$ds^2=-\left(1-\frac{2M}{r}+\frac{Q^2}{r^2}\right)dt^2+\left(1-\frac{2M}{r}+\frac{Q^2}{r^2}\right)^{-1}dr^2+r^2\,d\Omega^2. \tag{9}$$

However, the transition between the black hole and naked singularity for the GMGHS solutions occurs at $Q^2=2e^{2\phi_0}M^2$ whereas the same occurs at $Q^2=M^2$ for the RN solutions. Another important difference is that there is no inner horizon in the GMGHS class of solutions. The Hawking temperature of the GMGHS black hole is

$$T_H=\frac{1}{8\pi Me^{\phi_0}}, \tag{10}$$

which is independent of charge. This is again different from the RN black hole for which the Hawking temperature is given by

$$T_H=\frac{\sqrt{M^2-Q^2}}{2\pi(M+\sqrt{M^2-Q^2})^2}. \tag{11}$$

All these facts show that the charged black hole solutions of string theory and general relativity are qualitatively different.

## III. LENSING DUE TO THE CHARGED STRING BLACK HOLE

We consider the lens geometry as follows. A light ray from a source $(S)$ is deflected by a lens $(L)$ of mass $M$ and reaches an observer $(O)$. The background spacetime is taken asymptotically flat, and both the source and the observer are placed in the flat space-time. The line joining the lens and the observer $(OL)$ is taken as the optic axis for this configuration. $\beta$ and $\theta$ are the angular position of the source and the image with respect to the optic axis, respectively. The distances between observer and lens, lens and source, and observer and source are $D_{OL}$, $D_{LS}$, and $D_{OS}$, respectively (all distances are expressed in terms of the Schwarzschild radius $r_s=2M$, where $M$ is the mass of the lens). The position of the source and the image are related through the so called lens equation [12]

$$\tan\theta-\tan\beta=\frac{D_{LS}}{D_{OS}}[\tan\theta+\tan(\alpha-\theta)], \tag{12}$$

where $\alpha$ is the deflection angle. For positive $\beta$, the above relation gives images only on the same side $(\theta>0)$ of the source. Images on the other side can be obtained by taking negative values of $\beta$. The first and main step in getting the image positions is to calculate the deflection angle.

For the GMGHS space-time the deflection angle as a function of closest approach $x_0$ $(x_0=r_0/2M)$ can be expressed as

$$\alpha(x_0)=I(x_0)-\pi, \tag{13}$$

where

$$I(x_0)=2\int_{x_0}^{\infty}\frac{dx/x}{\sqrt{(x/x_0)^2(1-1/x_0)(1-\xi/x)^2(1-\xi/x_0)^{-1}-(1-\xi/x)(1-1/x)}}, \tag{14}$$

$x=r/2M$, and $\xi=Q^2e^{-2\phi_0}/2M^2$. The relation between the impact parameter (the perpendicular distance from the lens to the tangent to the null geodesic of the source) and the distance of closest approach can be obtained from the conservation of the angular momentum of the scattering process, and it is given by

$$b(x_0)=x_0\sqrt{\frac{1-\xi/x_0}{1-1/x_0}}. \tag{15}$$

With the decrease of the closest approach $x_0$, the deflection angle will increase, and for a certain value of $x_0$ the deflection angle will become $2\pi$, so that the light ray will make a complete loop around the lens. If $x_0$ decreases further, the light ray will wind several times around the lens before reaching the observer, and finally, when $x_0$ is equal to the radius of the photon sphere ($x_{ps}$), the deflection angle will become unboundedly large and the incident photon will be captured by the black hole. The radius of the photon sphere for this black hole is given by

$$x_{ps}=\frac{\xi+3+\zeta}{4} \tag{16}$$

where

$$\zeta=\sqrt{\xi^2-10\xi+9}. \tag{17}$$

Expanding the integrand in Eq. (14) in powers of $1/x$ and taking up to the second order, we get the deflection angle as follows:

$$\alpha(r_0)=\frac{4M}{r_0}+\frac{4M^2}{r_0^2}\left(\frac{15\pi}{16}-1\right)-\frac{Q^2e^{-2\phi_0}}{r_0^2}\left(\frac{3\pi}{4}-2\right)+\frac{Q^4e^{-4\phi_0}}{M^2r_0^2}\left(2-\frac{\pi}{16}\right)+O\left(\frac{1}{r_0^3}\right). \tag{18}$$

One trivially recovers the Schwarzschild deflection angle by setting $Q=0$ in the above equation. This deflection angle is larger than that for the RN space-time, which is given by[1]

$$\alpha(r_0)=\frac{4M}{r_0}+\frac{4M^2}{r_0^2}\left(\frac{15\pi}{16}-1\right)-\frac{3\pi}{4}\frac{Q^2}{r_0^2}+O\left(\frac{1}{r_0^3}\right). \tag{19}$$

[1]The coefficient of $Q^2/r_0^2$ in Eq. (55) of [11] contains an additional term, which occurred due to a typing error in the calculation [14].

The term proportional to $Q^4/r_0^2$ is totally absent in the deflection angle for the RN black hole [11] but this term is obviously very small as long as $Q$ is small.

To evaluate the integral (14) close to its divergence, the divergent integral will be split into two parts to separate out the divergent $[I_D(x_0)]$ and the regular parts $[I_R(x_0)]$. Then both of them will be expanded around $x_0=x_{ps}$ and will be approximated by the leading terms. But first the integrand of Eq. (14) is expressed as a function of a new convenient variable $z$, which is defined by

$$z=1-\frac{x_0}{x}, \tag{20}$$

so that

$$I(x_0)=\int_0^1 R(z,x_0)f(z,x_0)dz, \tag{21}$$

where

$$R(z,x_0)=\frac{2\sqrt{1-\xi/x_0}}{1-\xi/x_0+z\xi/x_0}, \tag{22}$$

$$f(z,x_0)=\left[1-\frac{1}{x_0}-\left(1-\frac{1}{x_0}+\frac{z}{x_0}\right)(1-z)^2\times\left(1+\frac{z\xi}{x_0-\xi}\right)^{-1}\right]^{-1/2}. \tag{23}$$

The integral (21) is then split into two parts:

$$I(x_0)=I_D(x_0)+I_R(x_0), \tag{24}$$

where

$$I_D(x_0)=\int_0^1 R(0,x_{ps})f_0(z,x_0)dz \tag{25}$$

includes the divergence and

$$I_R(x_0)=\int_0^1 g(z,x_0)dz \tag{26}$$

is regular in $z$ and $x_0$. The function $f_0(z,x_0)$ is the expansion of the argument of the square root in the divergent function $f(z,x_0)$ up to the second order in $z$:

$$f_0(z,x_0)=\frac{1}{\sqrt{p(x_0)z+q(x_0)z^2}}, \tag{27}$$

where

$$p(x_0)=\frac{\xi(x_0-2)+x_0(3-2x_0)}{x_0(\xi-x_0)}, \quad (28)$$

$$q(x_0)=\frac{\xi^2-3\xi x_0-(x_0-3)x_0^2}{x_0(\xi-x_0)^2} \quad (29)$$

and the function $g(z,x_0)$ is simply the difference between the original integrand and the divergent integrand:

$$g(z,x_0)=R(z,x_0)f(z,x_0)-R(0,x_{ps})f_0(z,x_0). \quad (30)$$

The leading order term of the divergent integral $I_D(x_0)$ is given by

$$I_D(x_0)=-u_D\log\left(\frac{x_0}{x_{ps}}-1\right)+v_D+O(x_0-x_{ps}), \quad (31)$$

where

$$u_D=\frac{R(0,x_{ps})}{\sqrt{q(x_{ps})}}, \quad (32)$$

$$v_D=u_D\log 2, \quad (33)$$

and

$$q(x_{ps})=\frac{4(1-\xi)[\xi^2+\xi(-12+\zeta)+9(3+\zeta)]}{(3-3\xi+\zeta)^2(3+\xi+\zeta)}. \quad (34)$$

Although $I_R(x_0)$ is regular at $z=0$ and $x_0=x_{ps}$, it is difficult to solve exactly. Expanding $I_R(x_0)$ in power of $x_0-x_{ps}$ and considering only the first expansion term (the zeroth order term), one gets

$$I_R(x_0)=\int_0^1 g(z,x_{ps})dz+O(x_0-x_{ps}). \quad (35)$$

This integral can be evaluated exactly, and the regular term in the deflection angle becomes

$$v_R=I_R(x_{ps})=h\left[\log(4\sqrt{a_0})-\log\left(\frac{2a_0+a_1}{\sqrt{a_0}}+2\sqrt{a_0+a_1+a_2}\right)\right], \quad (36)$$

where

$$a_0=2(1-\xi)[\xi^2+\xi(-12+\zeta)+9(3+\zeta)], \quad (37)$$

$$a_1=4[\xi^3+\xi^2(-15+\zeta)+\xi(23+6\zeta)-3(3+\zeta)], \quad (38)$$

$$a_2=24\xi^2-8\xi(3+\zeta), \quad (39)$$

and

$$h=\frac{\sqrt{3-3\xi+\zeta}(3+\xi+\zeta)}{\sqrt{a_0/2}}. \quad (40)$$

Thus the final expression for the strong-field limit of the deflection angle becomes [13]

$$\alpha(\theta)\simeq -u\log\left(\frac{\theta D_{OL}}{b(x_{ps})}-1\right)+v+O(b-b(x_{ps})), \quad (41)$$

where

$$u=\frac{u_D}{2}=\frac{h}{2}, \quad (42)$$

$$\begin{aligned} v&=-\pi+v_R+\frac{h}{2}\log\frac{2q(x_{ps})}{1-1/x_{ps}}\\ &=-\pi+\frac{h}{2}\left\{2\left[\log(4\sqrt{a_0})-\log\left(\frac{2a_0+a_1}{\sqrt{a_0}}+2\sqrt{a_0+a_1+a_2}\right)\right]+\log\frac{4a_0}{(3-3\xi+\zeta)^2(-1+\xi+\zeta)}\right\} \end{aligned} \quad (43)$$

and the impact parameter at the photon sphere $b(x_{ps})$ is

$$b(x_{ps})=\frac{1}{2\sqrt{2}}\sqrt{(9-\xi)\zeta+27-18\xi-\xi^2}. \quad (44)$$

Setting $\xi$ to 0, one obtains the Schwarzschild deflection angle (in the strong limit) from Eq. (41) using Eqs. (42), (43), and (44).

In contrast, for the RN solution the strong-field limit of the deflection angle can also be approximated by Eq. (41) but with the following coefficients [13]:

$$u=\frac{x_{ps}\sqrt{x_{ps}-2q^2}}{\sqrt{(3-x_{ps})x_{ps}^2-9q^2x_{ps}+8q^4}}, \quad (45)$$

$$v\simeq -\pi+0.9496-1.5939q^2+u\log\left[2(x_{ps}-q^2)^2\frac{(3-x_{ps})x_{ps}^2-9q^2x_{ps}+8q^4}{(x_{ps}-2q^2)^3(x_{ps}^2-x_{ps}+q^2)}\right], \quad (46)$$

$$b(x_{ps})=\frac{(3+\sqrt{9-32q^2})^2}{4\sqrt{2}\sqrt{3-8q^2+\sqrt{9-32q^2}}}, \quad (47)$$

where the radius of the photon sphere $(x_{ps})$ for this black hole is

$$x_{ps}=\frac{3}{4}\left(1+\sqrt{1-\frac{32q^2}{9}}\right). \quad (48)$$

Here $q=Q/2M$. In this case Eq. (35) cannot be evaluated exactly. Hence the first regular term of the deflection angle is approximated only up to the $q^2$ term. Thus the coefficients of the deflection angles for the RN and GMGHS black holes are different.

TABLE I. Estimates of the lensing observable for the central black hole of our galaxy.

| Observable | GMGHS | | | | Reissner-Nordström | | | |
|---|---|---|---|---|---|---|---|---|
| | $Q$ | | | | $Q$ | | | |
| | $0.1M$ | $0.2M$ | $0.4M$ | $0.8M$ | $0.1M$ | $0.2M$ | $0.4M$ | $0.8M$ |
| $\theta_\infty$ ($\mu$ arc sec) | 16.84 | 16.76 | 16.41 | 14.92 | 16.84 | 16.76 | 16.41 | 14.76 |
| $s$ ($\mu$ arc sec) | 0.0212 | 0.0216 | 0.0233 | 0.0326 | 0.0212 | 0.0216 | 0.0234 | 0.038 |
| $r$ | 6.81 | 6.79 | 6.70 | 6.27 | 6.81 | 6.79 | 6.69 | 6.07 |

Once the deflection angle is known, the positions of the images can be obtained from Eq. (12). In the strong-field regime and when the source, lens, and observer are highly aligned, the lens equation becomes [10]

$$\beta = \theta - \frac{D_{LS}}{D_{OS}} \Delta\alpha_n , \tag{49}$$

where $\Delta\alpha_n = \alpha - 2n\pi$ is the offset of the deflection angle $\alpha$ and $n$ is an integer. If $\theta_n^0$ are the image positions corresponding to $\alpha = 2n\pi$, we have, from Eq. (41),

$$\theta_n^0 = \frac{b(x_{ps})}{D_{OL}}(1+e_n), \tag{50}$$

where

$$e_n = e^{(v-2n\pi)/u}, \tag{51}$$

and thus the position of the $n$th relativistic image can be approximated by [13]

$$\theta_n = \theta_n^0 + \frac{b(x_{ps}) e_n D_{OS}}{u D_{LS} D_{OL}} (\beta - \theta_n^0). \tag{52}$$

The magnification of the $n$th relativistic image is given by (approximating the position of the images by $\theta_n^0$)

$$\mu_n = \frac{1}{(\beta/\theta)\partial\beta/\partial\theta} \simeq e_n \frac{b(x_{ps})^2 (1+e_n) D_{OS}}{u\beta D_{LS} D_{OL}^2}. \tag{53}$$

In the simplest situation, if only the outermost image can be resolved as a single image, then its angular separation from the remaining bunch of relativistic images is

$$s = \theta_1 - \theta_\infty , \tag{54}$$

where $\theta_\infty$ is the angular position of a set of relativistic images in the limit $n \to \infty$. If $r$ denotes the ratio of the flux from the outermost relativistic image to those from the remaining relativistic images, then

$$r \simeq e^{2\pi/u}. \tag{55}$$

Since the deflection angle is already known, the strong-lensing parameters, viz., the positions of the relativistic images, the angular separation between the outermost relativistic image and the remaining relativistic images, and their flux ratio, readily follow from Eqs. (50), (54), and (55). By measuring these parameters one should be able to identify the nature of the lensing black hole [13].

## IV. LENSING BY THE SUPERMASSIVE GALACTIC CENTER

Nuclear stellar dynamics indicates the existence of supermassive central black holes in our galaxy as well as in many other galaxies [7]. Considering the supermassive galactic center as a GMGHS black hole, we estimate the observables of strong lensing. The present example is intended only to estimate numerical values of the lensing parameters so that we get some idea of the string effect in a strong-lens observation.

The mass of the central object of our galaxy is estimated as $2\times10^6$ of the solar mass and its distance is around 8.5 kpc. Therefore $D_{OL} \sim 3.18\times10^{10}$. Taking the source distance $D_{OS} = 2D_{OL}$, the angular position of the relativistic images ($\theta_\infty$), the angular separation of the outermost relativistic image with the remaining bunch of relativistic images ($s$), and the relative magnification of the outermost relativistic image with respect to the other relativistic images ($r$) are estimated and are given in Table I. Here we have taken $\phi_0 = 0$ (in the string $\sigma$ model $e^{\phi_0}$ can be identified with the parameter $\alpha$ of the Eddington-Robertson metric, and the empirical definition of the mass thus leads to the choice $e^{\phi_0} = 1$). The same observable parameters when the lens is a Reissner-Nordström black hole instead of a GMGHS black hole are also given in Table I for comparison. It is clear from Table I that for small $Q$ the observational predictions of the GMGHS and RN black holes are the same within the given accuracy. For large $Q$, however, the lensing parameters are different for these two classes of black holes. But this is mainly because the first regular term in the deflection angle for the RN lens has been approximated up to the second order in $Q$, whereas the same term for the GMGHS black hole is exact.

On the other hand, if we consider a black hole having mass equal to the solar mass and charge $Q$ situated at the galactic halo with distance about 4 kpc as the lens, and a star in the galactic bulge (distance $\sim 8$ kpc) as the source, then $D_{OL}$ will be $\sim 4.19\times10^{16}$. In that case, for $Q = 0.1$ the positions of the relativistic images will be $\theta_\infty = 1.28\times10^{-5}$ $\mu$ arc sec and $s = 1.61\times10^{-5}$ $\mu$ arc sec. The relative magnification $r$ will remain the same as in the case of lensing by the galactic center (Table I).

## V. DISCUSSION

When the curvature is small compared to the Planck scale, all vacuum solutions of the Einstein field equations are approximate solutions of string theory. But in the regime of strong curvature the solutions of the two theories differ fundamentally. The same is also true when additional matter fields, such as the Maxwell field, are present. As a result, the charged black hole solutions of general relativity are not even considered as an approximate solution of string theory. Thus, it is expected that there will be some distinctive observational features of these two theories, particularly in the strong-gravity scenario and when matter fields are involved. To unveil such features we studied the gravitational lensing due to the charged black hole of heterotic string theory and calculated different strong-lensing parameters, such as the angular positions of the relativistic images, angular separation between the outermost relativistic image and the rest of the images, and also their relative magnification.

Modeling the massive compact object at the center of the galaxy as a GMGHS black hole, we estimated the numerical values of different strong-lensing parameters. When compared with the corresponding lensing observable due to the Reissner-Nordström black hole, it is found that against the expectation there is no significant string effect present on the observable parameters. This means that, although the GMGHS and the RN black hole solutions look quite different, in respect of the observational effects, particularly in the case of strong-field lensing, these two black holes are nearly the same.

It has already been realized that observation of relativistic images is very difficult [9]. To observe the relativistic images, the resolution of the detecting telescope needs to be of the order of $\mu$ arc sec or even better, whereas the resolution achieved so far is only of the order of m arc sec. More importantly, the relativistic images are highly demagnified. But if future experiments can attain 0.01 $\mu$ arc sec resolution and if ever such highly demagnified relativistic images are detected, even then lensing observations will not be able to distinguish the string theory from general relativity.

[1] J. Scherk and J.W. Schwarz, Nucl. Phys. **B81**, 118 (1974); D.J. Gross and E. Witten, *ibid.* **B277**, 1 (1986); R. Myers, *ibid.* **B289**, 701 (1987); C.G. Callan, R.C. Myers, and M.J. Perry, *ibid.* **B311**, 673 (1988), and references therein.
[2] G.W. Gibbons, Nucl. Phys. **B207**, 337 (1982); G.W. Gibbons and K. Maeda, *ibid.* **B298**, 741 (1988); D. Garfinkle, G.T. Horowitz, and A. Strominger, Phys. Rev. D **43**, 3140 (1991); **45**, 3888(E) (1992).
[3] J.D. Gegenberg, Gen. Relativ. Gravit. **21**, 155 (1989).
[4] A. Sen, Phys. Rev. Lett. **69**, 1006 (1992).
[5] A. Bhadra, T.B. Nayak, and K.K. Nandi, Phys. Lett. A **295**, 1 (2002).
[6] P. Schneider, J. Ehlers, and E.E. Falco, *Gravitational Lenses* (Springer-Verlag, Berlin, 1992).
[7] D. Richstone *et al.*, Nature (London) **395**, A14 (1998); J. Magorrian *et al.*, Astron. J. **115**, 2285 (1998).
[8] S. Frittelli, T.P. Kling, and E.T. Newman, Phys. Rev. D **61**, 064021 (2000).
[9] K.S. Virbhadra and G.F.R. Ellis, Phys. Rev. D **62**, 084003 (2000).
[10] V. Bozza, S. Capozziello, G. Iovane, and G. Scarpetta, Gen. Relativ. Gravit. **33**, 1535 (2001).
[11] E.F. Eiroa, G.E. Romero, and D.F. Torres, Phys. Rev. D **66**, 024010 (2002).
[12] K.S. Virbhadra and G.F.R. Ellis, Phys. Rev. D **65**, 103004 (2002).
[13] V. Bozza, Phys. Rev. D **66**, 103001 (2002).
[14] E.F. Eiroa, G.E. Romero, and D.F. Torres (private communication).